\documentclass[reprint,superscriptaddress,aps,prl]{revtex4-1}
\usepackage{amsmath,epsfig,amssymb,subfigure,bm,dsfont}
\usepackage{lipsum} 
\usepackage{color}
\usepackage{graphicx}
\usepackage{dcolumn}
\usepackage{upgreek}
\usepackage{bm}
\usepackage{bbold}
\usepackage{amsfonts,color}
\usepackage{amsmath}
\usepackage{textcomp}
\usepackage{siunitx} 
\usepackage{gensymb}
\usepackage[unicode=true, breaklinks=false, pdfborder={0 0 1}, backref=false, colorlinks=true, linkcolor=blue, urlcolor=blue, citecolor=blue]{hyperref}

\newcounter{lastnote}

\begin{document}
\title{Observation and Control of Chiral Spin Frustration in BiYIG Thin Films}

\author{Jinlong Wang}
\thanks{These authors contributed equally to this work.}
\affiliation{%
Fert Beijing Institute, MIIT Key Laboratory of Spintronics, School of Integrated Circuit Science and Engineering, Beihang University, Beijing 100191, China
}%
\affiliation{%
International Quantum Academy, Shenzhen 518048, China
}%

\author{Hanchen Wang}
\thanks{These authors contributed equally to this work.}
\affiliation{%
Department of Materials, ETH Zurich, Zurich 8093, Switzerland
}%

\author{Zhewen Xu}
\thanks{These authors contributed equally to this work.}
\affiliation{%
Department of Physics, ETH Zurich, Zurich 8093, Switzerland
}%

\author{Artim L. Bassant}
\thanks{These authors contributed equally to this work.}
\affiliation{%
Institute for Theoretical Physics, Utrecht University, Utrecht 3584CC, The Netherlands}%

\author{Junfeng~Hu}
\affiliation{%
International Quantum Academy, Shenzhen 518048, China
}

\author{Wenjie Song}
\affiliation{%
International Quantum Academy, Shenzhen 518048, China
}%

\author{Chaozhong Li}
\affiliation{%
Key Laboratory for Magnetism and Magnetic Materials of the Ministry of Education, Lanzhou University, Lanzhou 730000, China
}%

\author{Xiangrui Meng}
\affiliation{%
 Interdisciplinary Institute of Light-Element Quantum Materials and Research Center for Light-Element Advanced Materials and International Center for Quantum Materials and Electron Microscopy Laboratory, School of Physics, Peking University, Beijing, China}%

\author{Mengqi Zhao}
\affiliation{%
State Key Laboratory of Low-Dimensional Quantum Physics and Department of Physics, Tsinghua University, Beijing 100084, China}%

\author{Song Liu}
\affiliation{%
International Quantum Academy, Shenzhen 518048, China
}%

\author{Guozhi~Chai}
\affiliation{%
Key Laboratory for Magnetism and Magnetic Materials of the Ministry of Education, Lanzhou University, Lanzhou 730000, China
}%

\author{Peng Gao}
\affiliation{%
 Interdisciplinary Institute of Light-Element Quantum Materials and Research Center for Light-Element Advanced Materials and International Center for Quantum Materials and Electron Microscopy Laboratory, School of Physics, Peking University, Beijing, China}%

\author{Wanjun Jiang}
\affiliation{%
State Key Laboratory of Low-Dimensional Quantum Physics and Department of Physics, Tsinghua University, Beijing 100084, China}%

\author{Desheng Xue}
\affiliation{%
Key Laboratory for Magnetism and Magnetic Materials of the Ministry of Education, Lanzhou University, Lanzhou 730000, China
}%

\author{Dapeng Yu}
\affiliation{%
International Quantum Academy, Shenzhen 518048, China
}

\author{William Legrand}
\affiliation{%
Department of Materials, ETH Zurich, Zurich 8093, Switzerland
}
\affiliation{%
CNRS, Institut Néel, Université Grenoble Alpes, Grenoble 38042, France}

\author{Christian~L.~Degen}
\affiliation{%
Department of Physics, ETH Zurich, Zurich 8093, Switzerland
}
\affiliation{%
Quantum Center, ETH Zurich, Zurich 8093, Switzerland}

\author{Rembert A. Duine}
\email{r.a.duine@uu.nl}
\affiliation{%
Institute for Theoretical Physics, Utrecht University, Utrecht 3584CC, The Netherlands}%
\affiliation{%
Department of Applied Physics, Eindhoven University of Technology, P.O. Box 513, 5600 MB Eindhoven, The Netherlands
}

\author{Pietro Gambardella}
\email{pietro.gambardella@mat.ethz.ch}
\affiliation{%
Department of Materials, ETH Zurich, Zurich 8093, Switzerland
}%

\author{Haiming Yu}
\email{haiming.yu@buaa.edu.cn}
\affiliation{%
Fert Beijing Institute, MIIT Key Laboratory of Spintronics, School of Integrated Circuit Science and Engineering, Beihang University, Beijing 100191, China
}%
\affiliation{%
International Quantum Academy, Shenzhen 518048, China
}%

\date{\today}

\begin{abstract} 
Chiral interactions within magnetic layers stabilize the formation of noncollinear spin textures, which can be leveraged to design devices with tailored magnetization dynamics. Here we introduce chiral spin frustration in which energetically degenerate magnetic states frustrate the Dzyaloshinskii-Moriya interaction. We demonstrate magnon-driven switching of the chirally frustrated spin states in Bi-substituted yttrium iron garnet thin films. These states are defined by an in-plane macro-spin neighbouring two out-of-plane spins on either side with opposing chirality. Using scanning nitrogen-vacancy magnetometry and spin pumping, we identified four degenerate frustrated states and achieved their controllable switching via magnon spin torque. Crucially, the switching is unidirectional, with selectivity determined by the incoming magnon direction. This mechanism provides a powerful approach to manipulate frustrated spin states with magnons. Chiral spin frustration unlocks the geometry constrains of conventional frustration, and therefore opens new horizons for frustrated magnetism, paving the way for energy-efficient spintronic devices based on frustration.

\end{abstract}

\maketitle
\begin{figure}
\includegraphics[width=88mm]{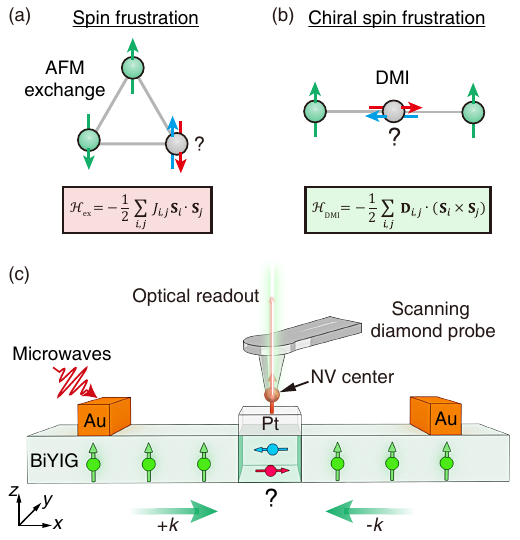}
\caption{(a) Conceptual sketch of spin frustration based on antiferromagnetic (AFM) exchange interactions between neighboring spins ($\bm{S}_{i}$ and $\bm{S}_{j}$). The AFM exchange constant $J_{i,j} < 0$. The Hamiltonian is determined by the dot product of neighboring spins. (b) Conceptual sketch of chiral spin frustration based on DMI between neighboring spins ($\bm{S}_{i}$ and $\bm{S}_{j}$). The DMI vector $\bm{D}_{i,j}$ points out of the paper plane. The Hamiltonian is determined by the cross product of neighboring spins. (c) Schematic of chiral spin frustration in a BiYIG thin film probed by nanoscale NV magnetometry. Magnons are excited by microwave antennas and injected from left ($+k$) and right ($-k$) are used to switch between different frustrated states. The magnetic field is applied in the out-of-plane direction.}
\label{fig1}
\end{figure}

Spin frustration, a phenomenon where competing magnetic interactions prevent simultaneous minimization of all local exchange energies, has been a central topic in magnetism and condensed matter physics~\cite{balents2010spin-1}. It often arises in antiferromagnetic systems with geometrical constraints, such as in the Ising model on a triangular~\cite{kurumaji2019skyrmion-2,sheng2025bose-3} (Fig.~\ref{fig1}(a)), square~\cite{wang2006artificial-4,gartside2022reconfigurable-5,lendinez2023nonlinear-6} or Kagome lattice~\cite{kang2020topological-7,bhat2020magnon-8} leading to degenerate ground states and exotic magnetic phases such as the spin nematic state~\cite{shannon2006nematic-9} and quantum spin liquids~\cite{broholm2020quantum-10}. Distinct from conventional bistable systems~\cite{peter-55,fert-56,Parkin-57}, frustrated states with low energy barriers~\cite{khanh2020nanometric-11,yang2020giant-12,xu2023frustration-13,prichard2024directly-14} provide a unique platform for magnetic switching at ultra-low powers and thus novel functionalities in spintronics. While conventional spin frustration based on collinear antiferromagnetic exchange interaction relies on noncollinear geometry (Fig.~\ref{fig1}(a)), chiral spin frustration based on noncollinear Dzyaloshinskii-Moriya interaction~\cite{halg2014quantum-15} (DMI) can simply be established in collinear spin chains (Fig.~\ref{fig1}(b)). To date, chiral spin frustration with multiple degenerate ground states has not been observed.

Recently, chiral spin coupling in ultrathin films of Co/Pt induced by the DMI has been utilized to realize synthetic antiferromagnets, field-free magnetization switching by spin-orbit torques, and domain wall logic devices~\cite{luo2019chirally-16,luo2020current-17}. Moreover, DMI-induced chiral spin textures have been used to tailor the emission and propagation of spin waves (magnons) in magnonic circuits~\cite{petti2022review-18,yu2021magnetic-19,zhang2025switchable-20}. Magnons can transfer spin angular momentum without an associated charge current, enabling low-power spintronic devices~\cite{kruglyak2010magnonics-21,lenk2011building-22,chumak2015magnon-23,pirro2021advances-24}. A particularly intriguing concept is magnon spin torque~\cite{slonczewski2010initiation-25,yan2011all-26}, where propagating spin waves exert torques on magnetization, enabling domain wall motion~\cite{han2019mutual-27,fan2023coherent-28} or magnetic switching~\cite{wang2019magnetization-29,baumgaertl2023reversal-30,wang2024deterministic-31,mucchietto2024magnon-32}. However, the manipulation of frustrated spin configurations by magnons has not been demonstrated.

In this work, we introduce frustration of the DMI, henceforth called chiral spin frustration, in Bi-substituted yttrium iron garnet (BiYIG) thin films~\cite{caretta2020relativistic-33}, where an in-plane (IP) macro-spin is non-collinearly coupled between two out-of-plane (OOP) macro-spins, as shown in Fig.~\ref{fig1}(b). Due to the presence of DMI, each OOP region favors aligning the IP macro-spin following the chirality set by minimizing the DMI energy. Consequently, the central IP macro-spin exhibits two energetically degenerate frustrated states, which were characterized by the single nitrogen-vacancy (NV) magnetometry~\cite{balasubramanian2008nanoscale-34} as shown in Fig.~\ref{fig1}(c). Chiral magnon switching behavior is demonstrated with the nonlocal spin pumping (SP) technique~\cite{kajiwara2010transmission-35,d2013inverse-36,wang2024broad-37}, where magnons driven from the left +$k$ (right -$k$) side can (cannot) switch the frustration state, and magnons driven from the right -$k$ (left +$k$) side can (cannot) switch it back. Finally, this observation is accounted for by a theoretical model considering the magnon spin torque influenced by the spin frustration due to the DMI.

Chirally frustrated spin states are experimentally realized in a 4 nm-thick BiYIG (Bi$_{0.8}$Y$_{2.2}$Fe$_5$O$_{12}$) film, grown on a Y$_3$Sc$_2$Ga$_3$O$_{12}$ substrate characterized by transmission electron microscopy, as shown in the Supplemental Material (SM)~\cite{SI-38}. The BiYIG thin film is grown with a compensated magnetic anisotropy near IP to OOP magnetic reorientation~\cite{SI-38} and engineered to exhibit low magnetic damping~\cite{soumah2018ultra-39}. The near compensation of effective magnetization is demonstrated by the ferromagnetic resonance~\cite{SI-38}. As shown in Fig.~\ref{fig1}(c), the film is capped partially with a Pt bar, known for its strong spin-orbit coupling, introducing an IP anisotropy on BiYIG~\cite{lee2020interfacial-40,lee2023large-41}. More details of the sample and devices are provided in SM~\cite{SI-38}. Importantly, the DMI of the BiYIG/Pt bilayer is measured with a DMI constant of $D=5.0 \pm 0.4\,\upmu\text{J}/\text{m}^2$ ($-1.8 \pm 0.2\,\upmu\text{J}/\text{m}^2$ without Pt capping) by Brillouin light scattering spectroscopy~\cite{SI-38,cortes2013influence-42,moon2013spin-43,nembach2015linear-44,kuepferling2023measuring-45}. The DMI constant of BiYIG/Pt is comparable to the value recently found in thulium iron garnet/Pt~\cite{husain2024field-46}. By applying a moderate OOP magnetic field, the uncapped regions can easily reach their magnetic saturation, whereas the Pt-capped region remains in-plane magnetized owing to the Rashba effect at the interface~\cite{lee2020interfacial-40}, enabling the formation of frustrated spin states associated with the DMI.

\begin{figure*}
\includegraphics[width=183mm]{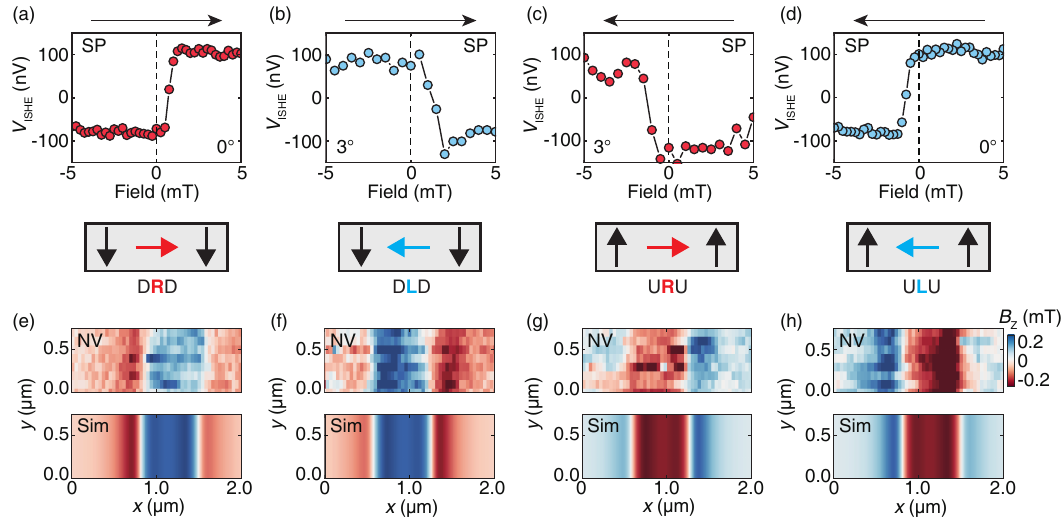}
\caption{(a) SP measurement at the excitation frequency of 0.35 GHz with the magnetic field applied perpendicular to the film ($0\degree$). The negative inverse spin-Hall voltage $V_\text{ISHE}$ indicates that the central macro-spin points to the right. Horizontal black arrows indicate the sweeping directions of the magnetic field. (b) SP measurement with the same conditions as (a), except for a tilt magnetic field angle of $3\degree$ towards the right side. The positive $V_\text{ISHE}$ indicates that the central macro-spin points to the left. (c) SP measurement with the same conditions as (b), except for a reversed field sweep direction. (d) SP measurement with the same conditions as (a), except for a reversed field sweep direction. (e)-(h), Four chiral spin frustration states imaged by scanning NV magnetometry showing different spatial profiles for the out-of-plane stray field ($B_\text{z}$), corresponding to simulation (Sim) results for frustration states of DRD (e), DLD (f), URU (g) and ULU (h), respectively.}  
\label{fig2}
\end{figure*}

To identify the frustrated states beneath the Pt, we employ both static (scanning NV microscopy) and dynamic (nonlocal SP) techniques to characterize the $x$-component of the magnetization $\mathbf{m}_x$. The device studied in this work is illustrated in Fig.~\ref{fig1}(c) (see SM~\cite{SI-38} for its microscope image). Nano-stripline (NSL) antennas~\cite{ciubotaru2016all-47} are fabricated 2.5 $\upmu$m apart from the Pt bar to excite magnons on both sides (The details of nonlocal SP and magnon switching can be found in SM~\cite{SI-38}). As magnons arrive at the Pt-covered region, spin currents with polarization collinear to the local magnetization are pumped into the Pt bar. Due to the inverse spin Hall effect (ISHE), spin currents are converted into charge currents leading to the detection of an ISHE voltage $V_\text{ISHE}$~\cite{d2013inverse-36,wang2024broad-37,saitoh2006conversion-48}. Figure~\ref{fig2}(a) shows $V_\text{ISHE}$ measured along the $y$-axis as a function of the calibrated OOP magnetic field ~\cite{SI-38} swept from negative to positive values. The negative magnetic field aligns the macrospins in the bare BiYIG region along the down direction, while the positive field aligns them toward the up direction. The microwave excitation is set at a frequency of 0.35 GHz and the power of -15 dBm. The data at different frequencies and full field sweep range are presented in SM~\cite{SI-38}. The sign of detected $V_\text{ISHE}$ reflects the magnetization orientation along the $x$-axis $m_x$ governed by $\mathbf{V}_{\mathrm{ISHE}}^{(y)} \propto \mathbf{j}_{s} \times \mathbf{m}_{x}$, where $\mathbf{j}_\text{s}$ denotes the spin current pumped into the Pt along the $z$-axis~\cite{saitoh2006conversion-48}. Thus, the positive and negative $V_\text{ISHE}$ indicate left (L) and right (R) $m_x$, respectively. In Fig.~\ref{fig2}(a), $V_\text{ISHE}$ stays negative when the OOP magnetic field is swept from negative to positive. This indicates that the ground state at zero field is DRD, denoting central IP spin pointing right and OOP spins on two sides pointing down. Interestingly, when the applied field is tilted by only about $3\degree$ (towards right from OOP), the ground state at zero field becomes DLD (Fig.~\ref{fig2}(b)). With the reversed magnetic field sweep, two other frustrated states of URU and ULU can be obtained at zero field as shown in Figs.~\ref{fig2}(c) and ~\ref{fig2}(d). All four states of DRD, DLD, URU and ULU can be stabilized at zero field (depending on different field hysteresis) and are therefore energetically degenerate ground states of the spin system (see phase diagram in SM~\cite{SI-38}).

To further characterize chirally frustrated states, we employ scanning NV magnetometry~\cite{velez2019high-49} to directly image the magnetization alignment of different frustrated spin states. The (111)-oriented diamond probe~\cite{SI-38,zhu2023multicone-50} is positioned above the device to detect the stray field in the OOP direction via Zeeman splitting~\cite{rohner2019111-51}. Due to the translational symmetry along $y$-axis, the detected stray field arises from the contributions from both the $x$- and $z$-components of the magnetization. Four frustrated spin states are probed by the nanoscale NV microcopy and presented in Figs.~\ref{fig2}(e)-(h). A quantitative spin Seebeck measurement indicates the coexistence of $\mathbf{m}_x$ and $\mathbf{m}_y$ components of the magnetization underneath Pt, with an estimated $75\degree$ angle with respect to the $x$-axis~\cite{SI-38}. Based on the NV-centre stray-field sensing formalism~\cite{van2015nanometre-52,zhewen}, we simulate the stray fields of four frustrated spin states with central IP macro-spin ($\mathbf{m}_x$) pointing either left or right and OOP macro-spins on two sides pointing either up or down, assuming a finite domain-wall width between the IP domain and OOP domain to be 160 nm~\cite{SI-38}. By directly comparing the experimental data from the NV magnetometry with the simulation results, we attribute the four observed frustrated spin states to be DLD, DRD, ULU and URU as presented in Figs.~\ref{fig2}(e)-(h). Notably, the actual frustrated state emerging at the zero-field ground state critically depends on the initial field orientation and field sweep direction (Figs.~\ref{fig2}(a)-(d)), and may vary with different magnetic field calibration in the NV and SP experiments. This observation suggests that all four frustrated states can be stabilized at zero field, and namely DLD, DRD, ULU and URU are degenerate ground states of the system~\cite{SI-38}. This feature is distinct with the frustrated states studied previously~\cite{luo2019chirally-16,vermeulen2024magnetization-53} that do not stabilize at zero field and show no degeneracy. The NV imaging results are complementary with the SP experiments that probes the orientation of the central macro-spin. As both techniques are insensitive to $\mathbf{m}_y$, the comprehensive three-dimensional spin texture is not fully resolved and requires further investigation that is beyond the scope of this work.

\begin{figure}[t]
\includegraphics[width=88mm]{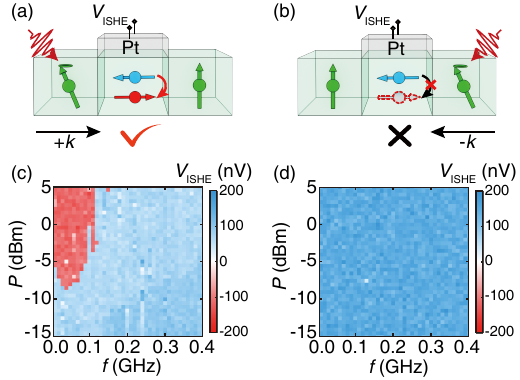}
\caption{(a) Sketch of effective switching by magnons injected from left ($+k$). (b) Sketch of ineffective switching by magnons injected from right ($-k$). (c) $V_\text{ISHE}$ detected as a function of excitation power ($P$) and frequency ($f$) in the configuration shown in (a) at zero magnetic field. (d) $V_\text{ISHE}$ detected as a function of $P$ and $f$ in the switching experiment of (b) at zero magnetic field.}
\label{fig3}
\end{figure}

In the following, we demonstrate the control of frustrated states by propagating magnons. Firstly, we initialize the system into the ULU state by sweeping external OOP field from 30 mT to 0 mT~\cite{SI-38} and then turn off the magnetic field. Subsequently, magnons are excited with microwave pulses ($\sim$1 ms) injected into the left NSL antenna (see Fig.~\ref{fig3}(a)). Magnons then propagate from left side into the frustrated region and exert magnon spin torque on the local magnetization. The SP measurement is employed to probe the local spin configuration~\cite{SI-38} at the microwave resonance of the IP region ($\sim$0.35 GHz) with a low power (-15 dBm). Fig.~\ref{fig3}(c) presents $V_\text{ISHE}$ detected after the magnon excitation as a function of microwave frequency and power. When excited at frequencies around the intrinsic resonance ($\sim$0.04 GHz) of the bare BiYIG film and with power above a threshold value of about -7 dBm (0.2 mW), the spin configuration at the frustrated region is switched from left to right (i.e., from ULU to URU) as indicated by the sign change of $V_\text{ISHE}$. The ultra-low power ($\sim$0.2 mW) required for the switching may arise from the soft magnons~\cite{bauer2023soft-54} excited in the fully compensated BiYIG film. However, the frustrated state does not switch even with the excitation directly at the IP region resonance with high powers up to 5 dBm. Notably, the switching from ULU to URU is successful only with magnon excitation from the left OOP region, but is ineffective if magnons are excited on the right side even with precisely the same conditions as presented in Figs.~\ref{fig3}(b) and \ref{fig3}(d) (Further details of this asymmetry switching are provided in the End Matter).  The unidirectional switching behavior may be considered as an inverse effect of the unidirectional spin-wave emitter in the presence of the DMI~\cite{DMI-pirro,DMI-hanchen}. If one needs to switch from URU back to ULU, the magnon excitation from the left side is ineffective and that from the right side instead can effectively trigger the switching~\cite{SI-38}. 

\begin{figure}[b]
\includegraphics[width=88mm]{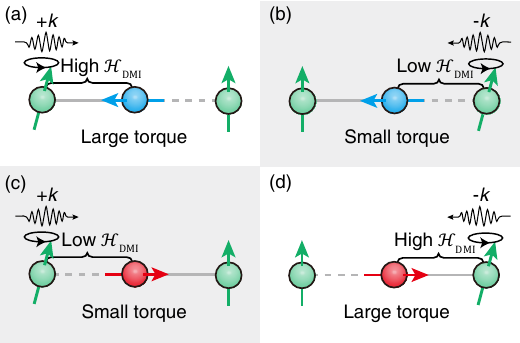}
\caption{(a) The frustrated state ULU where the left spin (U) and central spin (L) form high DMI energy $\mathcal{H}_\text{DMI}$, resulting in a large torque induced by incoming magnons from the left ($+k$). (b) The frustrated state ULU where the central spin (L) and right spin (U) form low DMI energy, resulting in a small torque induced by incoming magnons from the right ($-k$). (c) The frustrated state URU where the left spin (U) and central spin (R) form low DMI energy, resulting in a small torque induced by incoming magnons from the left ($+k$). (d) The frustrated state URU where the central spin (R) and right spin (U) form high DMI energy, resulting in a large torque induced by incoming magnons from the right ($-k$).}
\label{fig4}
\end{figure}

The unidirectional magnon switching is theoretically understood by considering a simplified model of the configuration as illustrated in Fig.~\ref{fig1}(b)~\cite{SI-38}. Fig.~\ref{fig4} presents four different situations for the magnon switching as experimentally studied in Fig.~\ref{fig3} and SM~\cite{SI-38}. In this theoretical model, the magnetization configuration is modelled by three single-domain parts, i.e., one macro-spin underneath the platinum pointing IP, coupled to two single macro-spins on either side pointing OOP. First, we consider the macro-spin underneath the Pt to point towards the left, i.e., with negative $\mathbf{m}_x$ (Fig.~\ref{fig4}(a)). Then the coupling via DMI to the left OOP spin is frustrated, which implies a high DMI energy $\mathcal{H}_\text{DMI}$ between the left and central spin. When exciting magnons from the left, i.e., when perturbing the left OOP spin, the variation of the free energy of the central spin will consequently be large~\cite{SI-38}, which in turn causes a large torque on the IP spin. This large torque ultimately switches the central spin from left to right. For the same situation, the coupling to the right OOP spin is not frustrated, therefore the interaction energy between the right OOP and central IP spin is at a minimum. Magnons injected from the right that perturb the right OOP spin will therefore exert little torque (Fig.~\ref{fig4}(b)). If the IP spin points to the right, i.e., with positive $m_x$, the same reasoning holds, which now leads to the conclusion that magnons coming in from the right more strongly perturbs the IP pointing spin and leads to switching (Figs.~\ref{fig4}(c) and~\ref{fig4}(d)). This explanation for the unidirectional switching is further supported by the calculations~\cite{SI-38}, which show that the magnon torque at zero applied field is given by 
\begin{equation}
\mathbf{\Gamma} = \frac{ 
  \left( \sqrt{1 - A^2} - 1 \right)}{ K } 
  \left( J \sqrt{K^2 - J^2} \pm DJ \right) 
\hat{y}.
\end{equation}
where $A$ is the magnon amplitude, $J$ the exchange energy, $K$ the in-plane anisotropy, and $D$ the DMI constant. The torque points in the $y$-direction, which causes the IP spin to rotate in-plane, and is large (+) when the coupling between the IP and OOP spins is frustrated and small (-) otherwise. Note that the torque not only results from frustration of the DMI but also from frustration of antiferromagnetic exchange~\cite{SI-38}. This shows that our results are more broadly applicable and may enable future magnonic devices based on frustration.

In summary, we investigated chiral spin frustration in magnetization-compensated BiYIG thin films partially capped with Pt, where the DMI and interfacial Rashba effect stabilize energetically degenerate frustrated states. These frustrated states are defined by an IP macro-spin in between two OOP spins with opposing chiral preference. Using two independent techniques, namely scanning NV magnetometry and nonlocal SP, we characterized four distinct frustrated spin states with consistent results. By magnon spin torque, we achieved unidirectional switching at low power ($\sim$0.2 mW) of these chirally frustrated states, that is dictated by the sign of the incoming magnon wavevector. Our work highlights the critical role of magnon spin torque in overcoming frustration barriers associated with the DMI, providing a novel mechanism to manipulate chiral spin frustration. Unlike bistable systems (e.g., spin valves~\cite{peter-55,fert-56} and magnetic tunnel junctions~\cite{Parkin-57}) requiring engineered anisotropies to maintain bistability, frustrated systems exhibit degenerate ground states emerging from intrinsic DMI. This frustration inherently softens the energy landscape, lowers the reversal barrier, and accelerates switching between degenerate states through low-energy magnon excitations. Moreover, our system bypasses the geometric constraints of conventional frustrated antiferromagnetic systems through noncollinear exchange interactions, enabling the coexistence of frustration and metastable states within a simple collinear high-symmetry magnet. Therefore, chiral spin frustration discovered in this work may open new opportunities for non-von Neumann architectures such as reservoir computing~\cite{Hu-58} and Hopfield networks~\cite{saccone2022direct-59}, with potential applications in artificial super-intelligence hardware.\\

\begin{acknowledgments}
$Acknowledgments$—The authors thank J. Li, L. Wu and R. Schlitz for their helpful discussions. We wish to acknowledge the support by the National Key Research and Development Program of China, Grants No. 2022YFA1402801; NSF China under Grants No. 12474104; China Scholarship Council (CSC) under Grant No. 202206020091; Swiss National Science Foundation (Grant Nos. 200020\_200465 and CRSII5\_205987); A.L.B. and R.A.D. thank the Dutch Research Council (NWO) for funding via the projects “Black holes on a chip” with project number OCENW.KLEIN.502 and “Fluid Spintronics” with project number VI.C.182.069. W.L. acknowledges the support of the ETH Zurich Postdoctoral Fellowship Program (Grant No. 21-1 FEL-48).
\end{acknowledgments}

\bibliographystyle{unsrt}

\onecolumngrid 
\section{End Matter}
\twocolumngrid 

\begin{figure}[h]
\includegraphics[width=88mm]{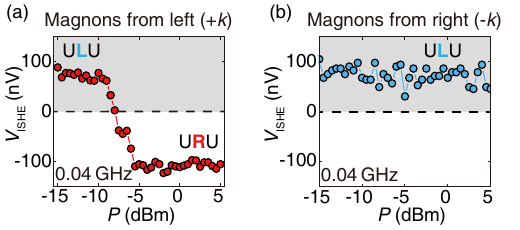}
\caption{(a) $V_\text{ISHE}$ detected after left-side magnons excitation at $f=0.04$ GHz with excitation power swept from negative to positive values. The $V_\text{ISHE}$ sign change indicates successful switching of frustrated spin states from ULU to URU. (b) $V_\text{ISHE}$ detected after right-side magnons excitation with the same excitation conditions as (a). The $V_\text{ISHE}$ sign remains unchanged indicating no switching.}
\label{fig5}
\end{figure}

$Appendix$—We here present details of the unidirectional magnon switching experiments. The magnetic system is initialized at the ULU state before being subjected to magnon excitation at 0.04~GHz (the resonance frequency of the adjacent OOP region). The excitation power is swept from -15~dBm to 5~dBm for magnons injected from either the left side [Fig.~\ref{fig5}(a)] or the right side [Fig.~\ref{fig5}(b)]. The results show that the ULU state can be efficiently switched to the URU state when the magnon spin current is injected from the left OOP region [Fig.~\ref{fig5}(a)]. When the power exceeds approximately -8~dBm, the frustrated state starts to switch progressively. As the power reaches around -5~dBm, the state is switched completely to URU, demonstrating a clear threshold behavior of the switching process. However, the switching is ineffective for magnons excited on the right side, even at the maximum applied power of 5~dBm [Fig.~\ref{fig5}(b)], the system remains at the ULU state. These results highlight the unidirectional switching behavior that the chirally frustrated spin states are only sensitive to incoming magnons from a favored direction.

\end{document}